# Anomaly Detection using Deep Autoencoders for in-situ Wastewater Systems Monitoring Data

**Stefania Russo[1], Andy Disch[1], Frank Blumensaat[1,2], Kris Villez[1]**

[1] Swiss Federal Institute of Aquatic Science and Technology, Eawag, Switzerland;
[2] Institute of Environmental Engineering, Chair of Urban Water Systems, ETH Zurich, Switzerland;
Email address: stefania.russo@eawag.ch; andy.disch@eawag.ch; frank.blumensaat@eawag.ch; kris.villez@eawag.ch

**Abstract:** Due to the growing amount of data from in-situ sensors in wastewater systems, it becomes necessary to automatically identify abnormal behaviours and ensure high data quality. This paper proposes an anomaly detection method based on a deep autoencoder for in-situ wastewater systems monitoring data. The autoencoder architecture is based on 1D Convolutional Neural Network (CNN) layers where the convolutions are performed over the inputs across the temporal axis of the data. Anomaly detection is then performed based on the reconstruction error of the decoding stage. The approach is validated on multivariate time series from in-sewer process monitoring data. We discuss the results and the challenge of labelling anomalies in complex time series. We suggest that our proposed approach can support the domain experts in the identification of anomalies.

**Keywords:** anomaly detection; machine learning; wastewater systems

## Introduction

In-situ sensors are increasingly used to monitor wastewater systems like urban drainage processes. This is thanks to the improvements in sensor technology and low-power data transmission techniques (Blumensaat *et al.*, 2017; Ebi *et al.*, 2019; Ruggaber *et al.*, 2007).

However, it is necessary to ensure high quality data, which plays an essential role in environmental monitoring since it can provide time and cost savings as well as immediate and actionable information (Racitiet *et al.*, 2012; Alferes et al., 2013; Leigh et al., 2019). Automated anomaly detection procedures are focused on automatically identifying unusual patterns or outliers that do not conform to the expected behaviour of the systems under investigation (Aggarwal, 2015). These procedures differ from manual anomaly detection, which is usually conducted by visual inspection of the raw or pre-processed data (Mourad and Bertrand-Krajewski, 2002). This requires expert handling and knowledge of the data and it is extensively time consuming, especially due to the growing amount and diversity of data.

Performing automated data validation for identifying abnormal behaviour of the deployed sensors is therefore critical. This is especially relevant within systems with high maintenance costs and a limited quality of service. In particular, suggested techniques and concepts to validate data from urban drainage systems (Alferes *et al.*, 2013, Branisavljevic, *et al.*, 2010) are not yet applicable to spatially distributed observations within such networks. Main challenges include the stochastic character of the natural driving force rainfall and partially confined signals, resulting from network-immanent flow constraints (throttles) and corresponding sewer overflows. Additionally, the nonlinear system response which depends on the amount of rainfall and uncontrolled boundary conditions exacerbates the application of the suggested anomaly detection methods.

Most of the methods in automated anomaly detection are based on statistical and machine learning algorithms (Hilland Minsker, 2010) and mainly rely on labelled data for training the learner algorithm. Unsupervised machine learning algorithms on the other hand, do not require labelled data, and have proven to be successful in several anomaly detection applications for environmental data (Ni et al., 2012; Ahmad et al., 2017; Laxhammar et al., 2009; Matteoli et al., 2010; Bezdek et al., 2011).

Between unsupervised methods, a particular family of architectures based on deep autoencoders have recently gained the attention of the research community (Domingos, 2015). Autoencoders are a type of deep neural networks that use non-linear dimensionality reduction to learn representations (encodings) of the data in an unsupervised manner (Hwang and Cho, 1999). Autoencoders consist of two stages: the encoding stage is used to reduce the dimensionality of the input data, while the decoding side aims at reconstructing the data back by minimizing the reconstruction error $L(x,\hat{x})$, which measures the difference between the original data $x$ and its reconstruction $\hat{x}$. In this way the network can "self-learn" the optimal parameters to represent the data via back-propagation.

Autoencoders can use different types of layers to learn meaningful representation of the data. Convolutional neural networks (CNNs) (LeCun et al., 1995) are still mostly used for object detection in image applications, but are starting to be considered a state-of-the-art method for classification tasks on time series and specifically for time series anomaly detection (Potes et al., 2016). In a recent study (Bai et al., 2018), temporal convolutional neural networks (TCN) have been shown to outperform recurrent neural networks (RNN) (Lipton et al., 2015) including their derivatives such as long-short term memory networks (LSTM) and gated recurrent networks (GRU), which are known to inherently handle time dynamics. However, the research on deep autoencoders based on CNN layers for anomaly detection on time series data is still in its infancy, with only few applications (Munir *et al.*, 2018; Wen and Zhang, 2018). This raises the question of whether the success of these architectures is restricted to specific applications or can be extended to other complex data.

In this study, we leverage on the advantages of using CNNs for representing time dynamics as stacked layers of deep autoencoders for unsupervised anomaly detection tasks, and prove the effectiveness of this combination over a case study presenting time series data from in-sewer process monitoring. In the following sections we present our network architecture and discuss the anomaly detection results on multivariate time series data from in-situ sensors in sewer collection systems. We then provide insights on broader applications and future directions.

**Material and Methods**

**Network architecture.** The architecture of an autoencoder presents an encoder stage which transforms the input into its compressed representation, and a decoder stage, which aims at reconstructing the original input from the lower dimensional representation. Therefore, the architecture presents a middle bottleneck, from which the reconstruction of the input data is implemented.

Our proposed architecture for the encoder stage uses 3 CNN layers where 1D convolutions are used with kernel size $k_1 = 8$, $k_2 = 6$, $k_3 = 4$, and filter maps $f_{maps1} = 64$, $f_{maps2} = 128$, $f_{maps3} = 256$. The convolution layers are followed by 2 fully connected layers and we use a ReLU activation after each layer (Nair and Hinton, 2010), and a dropout with rate = 0.1 for regularisation (Srivastava et al., 2014). This a standard configuration for CNN architectures, informed by recent research, and it is chosen to be simple, as in this work we are specifically aiming at validating the feasibility of our approach. The decoder stage is then configured to mirror the encoding process. The number of trainable parameters for this architecture is 5,174,147. During the end-to-end training we minimise the mean squared error loss over 100 epochs and a batch size = 4. The architecture is built on Keras with TensorFlow backend.

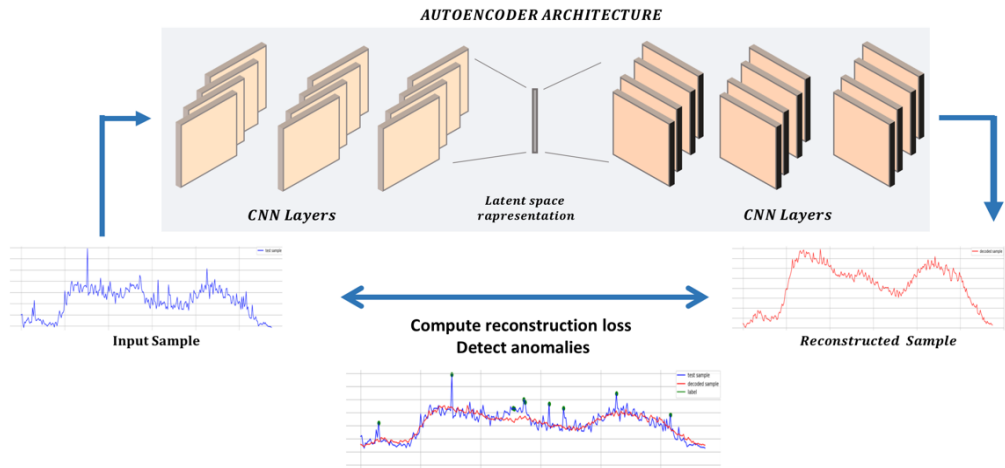

**Figure 1** Our proposed CNN autoencoder architecture for detecting anomalies in time series data.

**Case study**. Our case study presents data from in-sewer process monitoring campaign in the municipality of Fehraltorf, near Zurich in Switzerland, called the Urban Water Observatory initiative (*www.eawag.ch/uwo*). The catchment has a total area of 152 ha of which two thirds are drained through a combined sewer system. We use a dataset consisting of three sewer flow observations over a period of two full years (2017 and 2018) with a temporal resolution of 5 minutes. One observation is located downstream of the other two after a conjunction. Besides excluding data during wet weather periods, no further pre-processing was applied. We sample 119 time series segments from the dataset, each with length of T=288. This is done for the sake of considering a 24h data set.

Critical to quantitatively measure the results of our approach is the availability of a fully labelled data set of normal and abnormal behaviour. An annotation tool has therefore been developed as part of a software toolbox called "*AnomalyToolKit*," to perform the labelling procedure. Figure 2 shows an example of the labelling procedure conducted with this tool. This is applied to the time series data from in-sewer process monitoring, where we have asked an urban hydrology expert to inspect 24h time-series signals from the dataset. The expert performs the labelling considering his knowledge about the expected range of data values, the flow path topology and the characteristics of the deployed sensors.

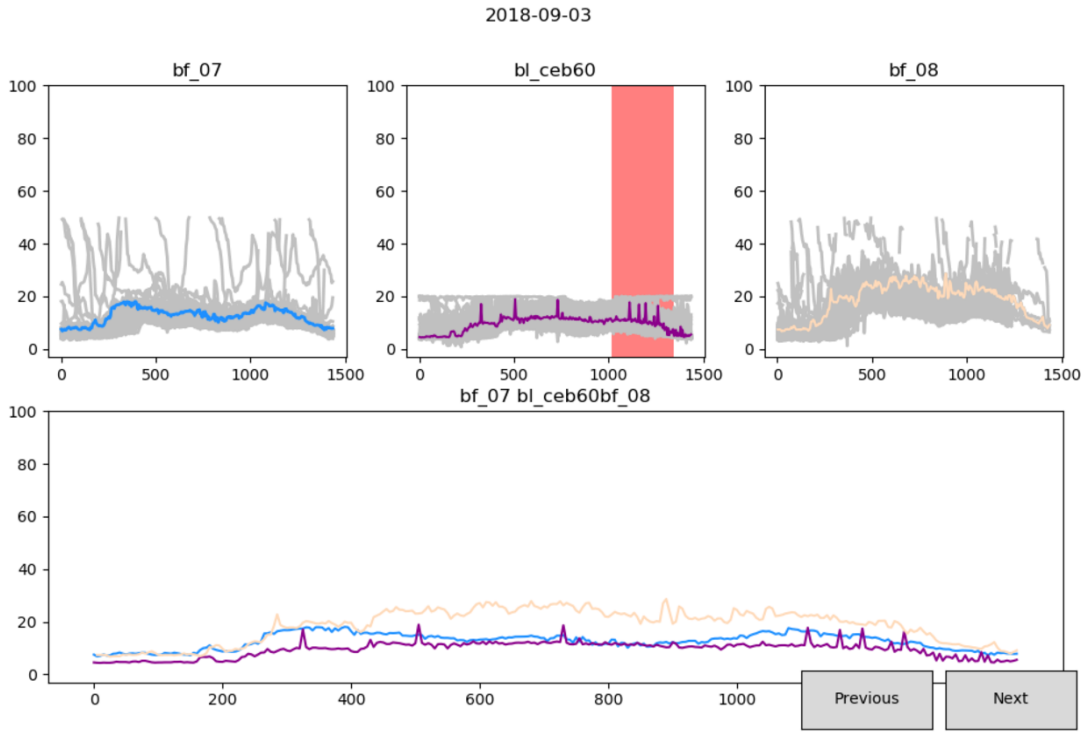

**Figure 2** Example of one daily time series which is manually labelled. Top panels show single sensors daily time series, while in the bottom panel it is plotted a combination of the three sensors. In grey, all the historical daily data are shown. The red span indicates the area selected by the urban hydrology expert where there is an anomaly. The expert can then move between the different time series by clicking on the buttons "Previous" and "Next".

These daily data sets are manually labelled, while inspecting all three flow time series at once to include the relation between the different signals and the corresponding "mental model" developed by the expert while labelling the data. The "mental model" refers to the dry weather pattern expected in a drainage system and therefore exploring the autocorrelation of the data within a daily cycle. We removed the labels of the first fifth of the data and added them again at the end to avoid systematic misjudgements and allow for a calibration phase of the "mental model". In order to carry out the time autoencoder training task, we first remove the labels and then split the sampled data in a way such that 20% is used for testing, and 80% for training and validation.

**Results and Discussion**

During model learning, the training and validation losses converge after only 55 epochs, as shown in Figure 3. The anomaly detection is then performed on unseen test data using the reconstruction error from the decoding stage as indication of anomaly. The identified anomalous points are compared to the previously labelled data to measure the performance of our approach. Our method presents a detection accuracy of 35%, measured as: $accuracy = \frac{\# \ of \ correctly \ identified \ anomalies}{total \ \# \ of \ anomalies}$. Considering the unsupervised scenario, the limited amount of training data samples and the choice for a traditional network architecture, this result demonstrates promise for unsupervised anomaly detection applications.

On the other hand, our method also shows a high number of false positives. After discussing with our labelling expert, we have noted that some of the anomalies identified by the autoencoder were correct, while the labelling expert did not label these as such due to the high complexity of the data and the pure amount of data to be labelled. A selection of these examples is shown in Figure 4. This makes our current results not yet quantifiable in terms of detection accuracy, but leaves a great opening for future research, in particular regarding the jointly usage of domain expertise and such unsupervised deep models. In fact, the labelling of anomalies in time series is not exact: a domain expert looks at the time series and based on his knowledge roughly highlights the time steps, which he deems anomalous. This does not always cover the true anomalous time series segments but overlaps with only parts of it. Furthermore, since this task is manual and takes up many labour hours, it is prone to error and inaccuracies. Our proposed approach can help the domain expert identifying anomalies in the data.

**Conclusions**

We have proposed an unsupervised anomaly detection method for in-situ wastewater systems monitoring data based on a deep autoencoder architecture with stacked 1D CNNs. The method detected a higher number of anomalies than expected. The hypothesis is that most of these anomalies were not initially identified by the domain expert due to the complexity of the data set. While this make an objective assessment challenging, our result suggests this tool to be useful, especially when used iteratively with the domain experts to identify anomalies in complex data sets.

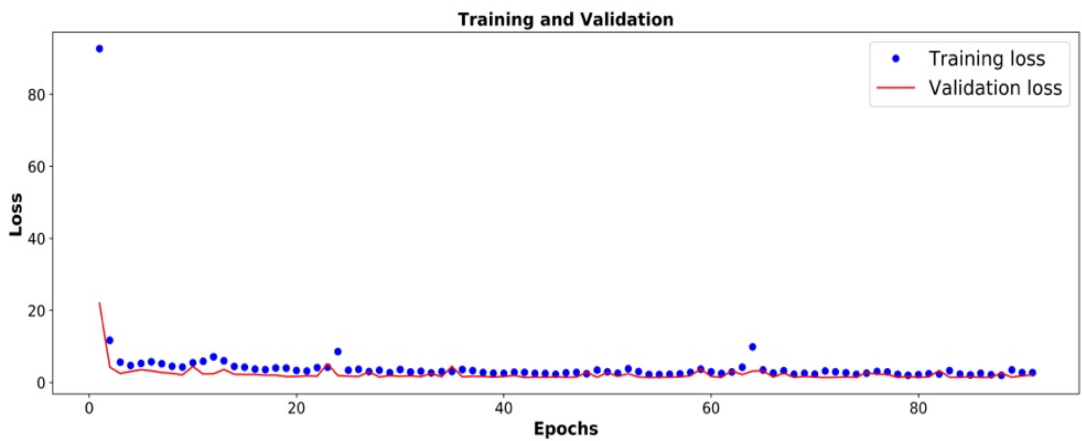

**Figure 3** During training, the training and validation losses converge after 55 epochs. We use early stopping to stop training once the validation accuracy starts to decrease in order to prevent overfitting.

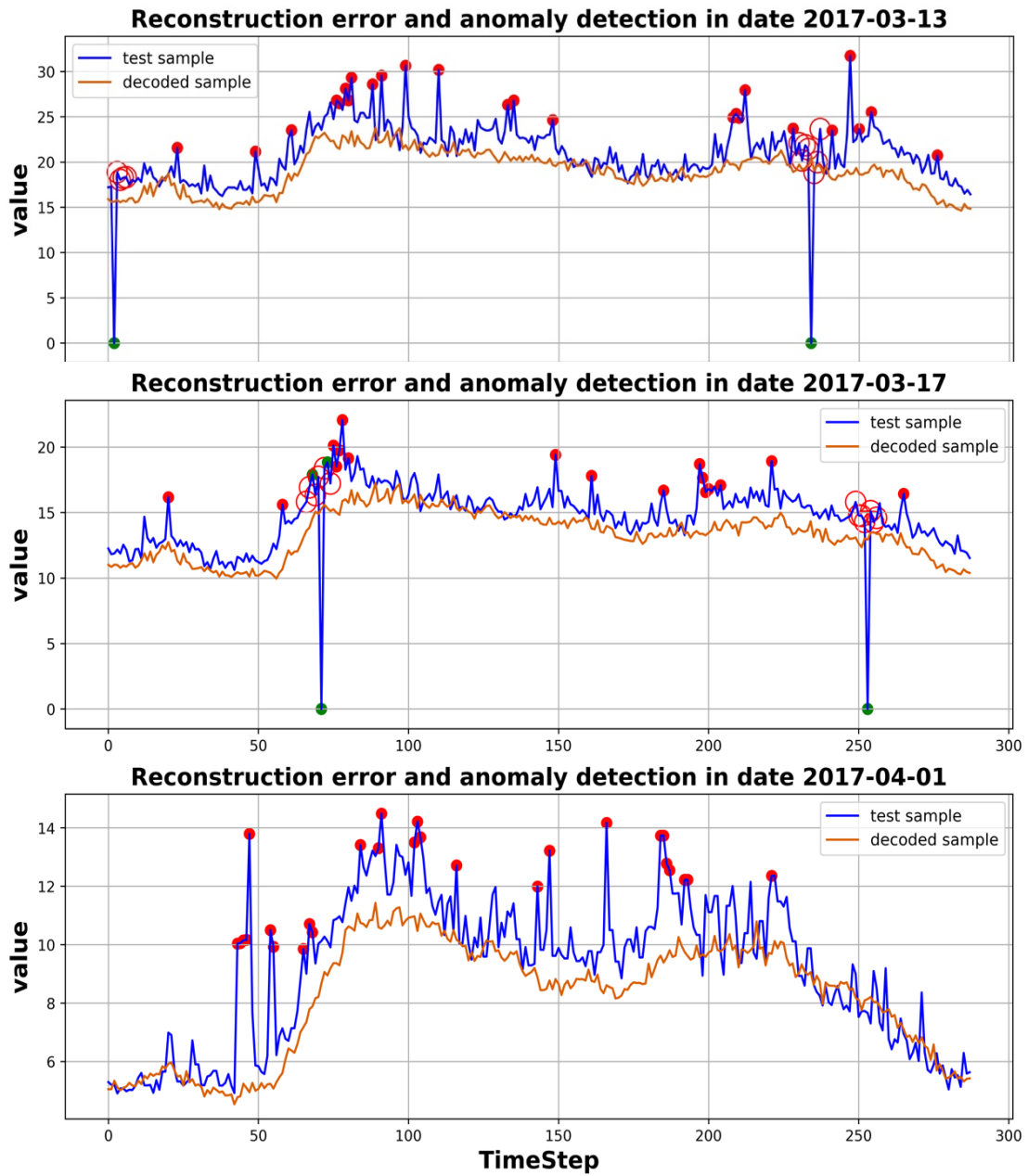

**Figure 4** Some examples of correctly and "incorrectly" identified anomalies. In the above figures, the true positives are indicated with a green filled circle, false positives with a red filled circle, false negatives with a red open circle. The current data labelling tool allows the expert to select time windows which present anomalies, however they possibly include normal data as well.

## Acknowledgements

This study has been made possible by the Eawag Discretionary Funds (grant number: 5221.00492.012.02, project:DF2018/ADASen).

# References

Aggarwal, C. C. (2015). Outlier analysis. In Data mining, 237–263.Springer.

Ahmad, S., Lavin, A., Purdy, S., and Agha, Z. (2017). Unsupervised real-time anomaly detection for streaming data. Neurocomputing, 262:134–147.

Alferes, J., Tik, S., Copp, J., and Vanrolleghem, P. A. (2013). Advanced monitoring of water systems using in situ measurement stations: data validation and fault detection. Water science and technology, 68(5):1022–1030.

Bai, S., Kolter, J. Z., and Koltun, V. (2018). An empirical evaluation of generic convolutional and recurrent networks for sequence modeling. arXiv preprintarXiv:1803.01271.

Bezdek, J. C., Rajasegarar, S., Moshtaghi, M., Leckie, C., Palaniswami, M., and Havens, T. C. (2011). Anomaly detection in environmental monitoring networks [application notes]. IEEE Computational Intelligence Magazine,6(2):52–58.

Branisavljevic, N., Prodanovic, D., and Pavlovic, D. (2010). "Automatic, semi-automatic and manual validation of urban drainage data." Water Science and Technology, 62(5), 1013-1021.

Blumensaat, F., Ebi, C., Dicht, S., Rieckermann, J., and Maurer, M. (2017). Langzeitüberwachung der Raum-Zeit-Dynamik in Entwässerungssystemen mittels Niedrigenergiefunk Ein Feldexperiment im Großmaßstab. Korrespondenz Abwasser, Abfall 7(64): 10.

Domingos, P. (2015). The master algorithm: How the quest for the ultimate learning machine will remake our world. Basic Books.

Ebi, C., Schaltegger, F., Rüst, A., & Blumensaat, F. (2019). Synchronous LoRa Mesh Network to Monitor Processes in Underground Infrastructure. IEEE Access, 7, 57663-57677. doi:10.1109/ACCESS.2019.2913985

Hill, D. J. and Minsker, B. S. (2010). Anomaly detection in streaming environ-mental sensor data: A data-driven modeling approach. Environmental Modelling & Software, 25(9):1014–1022.

LeCun, Y., Bengio, Y., et al. (1995). Convolutional networks for images, speech, and time series. The. handbook of brain theory and neural networks,3361(10):1995.

Leigh, C., Alsibai, O., Hyndman, R. J., Kandanaarachchi, S., King, O. C., Mc-Gree, J. M., Neelamraju, C., Strauss, J., Talagala, P. D., Turner, R. D., et al. (2019). A framework for automated anomaly detection in high frequency water-quality data from in situ sensors. Science of The Total Environment.

Lipton, Z. C., Kale, D. C., Elkan, C., and Wetzel, R. (2015). Learning to diagnose with lstm recurrent neural networks. arXiv preprint arXiv:1511.03677.

Matteoli, S., Diani, M., and Corsini, G. (2010). A tutorial overview of anomaly detection in hyperspectral images. IEEE Aerospace and Electronic Systems Magazine, 25(7):5–28.

Mourad, M. and Bertrand-Krajewski, J.-L. (2002). A method for automatic validation of long time series of data in urban hydrology. Water Science and Technology, 45(4-5):263–270.

Munir, M., Siddiqui, S. A., Dengel, A., and Ahmed, S. (2018). Deepant: A deep learning approach for unsupervised anomaly detection in time series. IEEE Access, 7, 1991-2005.

Nair, V. and Hinton, G. E. (2010). Rectified linear units improve restricted boltzmann machines. In Proceedings of the 27th international conference on ma-chine learning (ICML-10), pages 807–814.

Ni, J., Zhang, C., Ren, L., and Yang, S. X. (2012). Abrupt event monitoring for water environment system based on kpca and svm. IEEE Transactions on Instrumentation and Measurement, 61(4):980–989.

Potes, C., Parvaneh, S., Rahman, A., and Conroy, B. (2016). Ensemble of feature-based and deep learning-based classifiers for detection of abnormal heart sounds. In2016 Computing in Cardiology Conference (CinC), pages 621–624. IEEE.

Raciti, M., Cucurull, J., and Nadjm-Tehrani, S. (2012). Anomaly detection in water management systems. In Critical infrastructure protection, 98–119. Springer.

Ruggaber, T. P., Talley, J. W., and Montestruque, L. A. (2007). Using embedded sensor networks to monitor, control, and reduce cso events: A pilot study. Environmental Engineering Science, 24(2):172–182.

Srivastava, N., Hinton, G., Krizhevsky, A., Sutskever, I., and Salakhutdinov, R. (2014). Dropout: a simple way to prevent neural networks from overfitting. The Journal of Machine Learning Research, 15(1):1929–1958.

Wen, T. and Zhang, Z. (2018). Deep convolution neural network and autoencoders-based unsupervised feature learning of eeg signals. IEEE Access, 6:25399–25410.